\documentclass[11pt]{article}
\usepackage{graphicx}

\begin{document}

\def\xslash#1{{\rlap{$#1$}/}}
\def \p {\partial}
\def \dd {\psi_{u\bar dg}}
\def \ddp {\psi_{u\bar dgg}}
\def \pq {\psi_{u\bar d\bar uu}}
\def \jpsi {J/\psi}
\def \psip {\psi^\prime}
\def \to {\rightarrow}
\def\bfsig{\mbox{\boldmath$\sigma$}}
\def\DT{\mbox{\boldmath$\Delta_T $}}
\def\xit{\mbox{\boldmath$\xi_\perp $}}
\def \jpsi {J/\psi}
\def\bfej{\mbox{\boldmath$\varepsilon$}}
\def \t {\tilde}
\def\epn {\varepsilon}
\def \up {\uparrow}
\def \dn {\downarrow}
\def \da {\dagger}
\def \pn3 {\phi_{u\bar d g}}

\def \p4n {\phi_{u\bar d gg}}

\def \bx {\bar x}
\def \by {\bar y}


\begin{center}
{\Large\bf  Breakdown of QCD Factorization for P-Wave Quarkonium Production at Low Transverse Momentum }
\par\vskip20pt
J.P. Ma$^{1,2}$, J.X. Wang$^{3}$ and  S. Zhao$^{1}$     \\
{\small {\it
$^1$ State Key Laboratory of Theoretical Physics, Institute of Theoretical Physics, Academia Sinica,
P.O. Box 2735,
Beijing 100190, China\\
$^2$ Center for High-Energy Physics, Peking University, Beijing 100871, China  \\
$^3$ Institute of High Energy Physics, Academia Sinica, P.O. Box 918(4), Beijing 100049, China 
}} \\
\end{center}
\vskip 1cm
\begin{abstract}
Quarkonium production at low transverse momentum in hadron collisions can be used to extract 
Transverse-Momentum-Dependent(TMD) gluon distribution functions, if TMD factorization holds there. 
We show that TMD factorization for the case of  P-wave quarkonium with $J^{PC}=0^{++}, 2^{++}$  holds at one-loop level, 
but is violated beyond one-loop level.  TMD factorization for other P-wave quarkonium is also violated
already at one-loop level. 
\vskip 5mm
\noindent
\end{abstract}
\vskip 1cm
\par

\par

\par 
Collisions with hadrons provide important information about the interactions and inner structure of hadrons.
There are processes, where a small transverse momentum is involved. E.g., the lepton pair in Drell-Yan processes are produced in low transverse momentum. This type of processes is of particular interest. 
In general such a small transverse momentum is generated at least partly from the transverse motion 
of partons inside a hadron. Therefore, studies of the processes will provide information 
about transverse momentum distributions of parton in a hadron. 
\par 
In order to extract the distributions from experimental measurements, one needs to establish QCD
Transverse-Momentum-Dependent(TMD) factorizations to consistently separate nonperturbative- and perturbative 
effects in relevant processes. The nonperturbative effects are represented by TMD parton distribution functions and a soft factor. 
These quantities are defined with QCD operators. TMD factorization has been established for 
a number of processes like $e^+e^-$-annihilations\cite{CS},  Drell-Yan processes\cite{CSS,JMYP} and 
Semi-Inclusive Deeply Inelastic Scattering(SIDIS)\cite{JMY,CAM}. From Drell-Yan processes and SIDIS only TMD quark distribution functions can be extracted. Besides TMD quark distribution functions there exist 
TMD gluon 
distribution functions describing gluon contents of a hadron. 
Several processes like 
Higgs-production\cite{JMYG,SecG1}, quarkonium production \cite{BoPi}, two-photon production\cite{QSW} and 
the production of a quarkonium combined with a photon\cite{DLPS}, are suggested to determine TMD gluon 
distribution functions. In this letter we are interested in TMD factorization 
for P-wave quarkonium production at low transverse momentum. 

\par 
In the production of $^1S_0$ quarkonium at low transverse momentum, TMD factorization is explicitly examined at one-loop level in \cite{MWZ}, where the quarkonium will not interact with soft gluons at leading power.  Therefore, it is expected that the factorization holds beyond one-loop level. Here we will show that TMD factorization for 
$^3 P_0$ - and $^3 P_2$ quarkonium  holds at one-loop level. But it is violated beyond one-loop.
We will also discuss TMD factorization of $^1 P_1$ quarkonium.    
 
\par
We use the  light-cone coordinate system, in which a
vector $a^\mu$ is expressed as $a^\mu = (a^+, a^-, \vec a_\perp) =
((a^0+a^3)/\sqrt{2}, (a^0-a^3)/\sqrt{2}, a^1, a^2)$ and $a_\perp^2
=(a^1)^2+(a^2)^2$. $g_\perp^{\mu\nu}$ is the transverse part of the metric. Its nonzero elements are 
$g_\perp^{11}=g_\perp^{22}=-1$. The process we consider is:  
\begin{equation} 
   h_A (P_A) + h_B (P_B) \to \chi_{0,2}  (q) + X,
\label{proc}    
\end{equation}    
in the kinematic region with $q_\perp \ll M$, where $M$ is the mass of the quarkonium $\chi_0$ or $\chi_2$, 
i.e.,  $q^2 =M^2 =Q^2$ .   
We use  $\chi_J$ to denote the quarkonium  $\chi_{cJ}$ or $\chi_{bJ}$.
The momenta of initial hadrons are given by $ P_A^\mu \approx (P_A^+,0,0,0)$ and $P_B^\mu \approx (0,P_B^-, 0,0)$. We take the initial hadrons as unpolarized. 
\par 

It is noted that one can use collinear factorization for the process if the produced quarkonium has large 
transverse momentum, i.e., $q_\perp \gg \Lambda_{QCD}$. In this case the nonperturbative effects of initial hadrons are parameterized with standard parton distribution functions. The transverse momenta of partons from 
initial hadrons are neglected in comparison with large $q_\perp$. But, in the kinematic region with $q_\perp \sim \Lambda_{QCD} \ll Q$, the collinear factorization is not applicable. The transverse momenta of partons can not be neglected because they are at the order of $q_\perp$. In this region one may use 
TMD factorization. 

\par 
A quarkonium mainly consists a heavy quark $Q\bar Q$-pair. 
The heavy quark $Q$ or $\bar Q$ moves with a small velocity $v$ in the rest frame of the quarkonium. 
One can use nonrelativistic QCD to study a quarkonium.  To factorize the nonperturbative effects related to  a quarkonium in its production or its decay one can used NRQCD factorization suggested in \cite{nrqcd}. 
In this factorization, one makes an expansion in $v$ and the nonperturbative effects are represented by NRQCD 
matrix elements. At the leading order of $v$,  one needs to consider the production of a heavy quark $Q\bar Q$-pair  in color-singlet or color octet for a P-wave quarkonium.  The color-octet 
$Q\bar Q$-pair is in  $^3S_1$ -state and the singlet is in $^3P_J$-state. Hence, the production rate of a $^3P_J$-quarkonium can be written as a sum of two components at the leading order of $v$ in NRQCD factorization: 
 \begin{equation} 
  d\sigma (\chi_{J} ) = d\sigma (^3 P_J^{(1)}) + d\sigma (^3S_1^{(8)}). 
\label{TC} 
\end{equation}   
The first component denotes the contribution in which a $Q\bar Q$-pair is produced  in a color-single $^3P_J$-state 
and then the pair is transmitted into the quarkonium $\chi_J$.  The second component denotes the contribution in which a $Q\bar Q$-pair is produced  in a color-octet $^3S_1$-state 
and then the pair is transmitted into the quarkonium $\chi_J$.  The production of a heavy quark pair 
can be studied with perturbative QCD. The transmissions are nonperturbative and can be described with 
NRQCD matrix elements. We notice here that NRQCD factorization of the color-octet component 
can be violated at two-loop level and it can be restored by adding gauge links in NRQCD color-octet matrix 
elements\cite{NQS}.  
\par 

\par    
The production of a $Q\bar Q$-pair can be through different processes initiated by partons from initial hadrons. 
Because of high energy of initial hadrons, it is expected the production is initiated by gluons from hadrons in the 
initial state. At leading power or leading twist, the nonperturbative effects related to the initial hadrons 
are parametrized with TMD gluon distribution functions.  
We take 
$h_A$ to give the definitions. We first introduce the gauge link along the direction 
$u^\mu = (u^+, u^-,0,0)$:
\begin{equation}
{\mathcal L}_u (z,-\infty) = P \exp \left ( -i g_s \int^0_{-\infty}  d\lambda
     u\cdot G (\lambda u + z) \right ) , 
\end{equation}
where the gluon field is in the adjoint representation. At leading twist one can define two TMD gluon distributions
through the gluon density matrix\cite{JMYG,TMDGP}: 
\begin{eqnarray}
&& \frac{1}{x P^+} \int \frac{ d\xi^- d^2 \xi_\perp}{(2\pi)^3}
e^{ - i x \xi^- P^+_A + i \vec \xi_\perp \cdot \vec k_\perp}
 \langle h_A \vert \left ( G^{+\mu} (\xi ) {\mathcal L}_u (\xi,-\infty) \right )^a
            \left ( {\mathcal L}_u^\dagger (0,-\infty) G^{+\nu}(0) \right )^a \vert h_A \rangle
\nonumber\\
  &&   =-\frac{1}{2} g_\perp^{\mu\nu} f_{g/A} (x,k_\perp, \zeta^2_u,\mu)
   + \left (k_\perp^\mu k_\perp^\nu + \frac{1}{2}g_\perp^{\mu\nu} k_\perp^2 \right ) h_{g/A} (x,k_\perp, \zeta^2_u,\mu)  
\label{DEF} 
\end{eqnarray}
with $\xi^\mu =(0,\xi^-,\vec \xi_\perp)$. 
The definition is given in non-singular gauges. It is gauge invariant. In singular gauges, one needs to add 
gauge links along transverse direction at $\xi^-=-\infty$\cite{TMDJi}.  Because of the gauge links, the TMD gluon distributions also depend on the vector $u$ through the variable $\zeta^2_u = (2u\cdot P_A)^2/u^2$. 
In the definition the limit $u^+ \ll u^-$ is taken in the sense that one neglects all contributions suppressed 
by negative powers of $\zeta_u^2$. The TMD gluon distribution function of $h_B$ is defined in a similar way. 
Because $h_B$ moves in the $-$-direction, the used gauge link is along the direction $v^\mu =(v^+,v^-,0,0)$ 
with $v^+\gg v^-$. 
\par 
The above definition in Eq.(\ref{DEF}) is for an unpolarized hadron. There are two TMD gluon
distributions. The distribution $f_{g/A}$ describes unpolarized gluons in $h_A$ 
and can be related  to the
standard gluon distribution in collinear factorization, while 
the
distribution $h_{g/A}$ describes lineally polarized gluons in $h_A$.  
The phenomenology of $h_{g/A}$ has been 
recently studied \cite{SecG1,SecG2,SecG3}. In this work we will only consider the contributions with 
$f_{g/A}$.  
As showing in the studies of TMD factorization with TMD gluon distribution functions in \cite{JMYG,MWZ}, 
one needs a soft factor to factorize the effect of exchanges of soft gluons. The soft factor $\tilde S$  is defined as:
\begin{eqnarray}
\tilde S(\vec\ell_\perp,\mu,\rho) &&= \int\frac{d^2 b_\perp}{(2\pi)^2} e^{ i\vec b_\perp\cdot \vec\ell_\perp}
  S^{-1}(\vec b_\perp,\mu,\rho)
\nonumber\\
S(\vec b_\perp,\mu,\rho) &=& \frac{1}{N_c^2-1} \langle 0\vert {\rm Tr} \left  [   {\mathcal L}^\dagger_v (\vec b_\perp,-\infty)
  {\mathcal L}_u (\vec b_\perp,-\infty) {\mathcal L}_u^\dagger (\vec 0,-\infty){\mathcal L}_v (\vec 0,-\infty)   \right ] \vert 0\rangle.
\label{SoftS}
\end{eqnarray}
The defined TMD gluon distribution functions and the soft factor are nonperturbative ingredients in TMD factorizations in the mentioned processes for extracting TMD gluon distributions. 
The importance of TMD factorization is not only limited 
for exploring inner structure of initial hadrons, but also for resummation of large log terms 
in perturbative coefficient functions in collinear factorizations. Studies of the resummation 
in quarkonium production in kinematical regions of moderate transverse momenta have been carried out in 
\cite{BQW,SuYu1}

\par 
In general, a QCD  factorization, which is proven for a hadronic process, also holds if one replaces 
hadrons in the hadronic process with partons.  It  means that one can examine a factorization 
with corresponding partonic state. In our case, especially for showing violation of TMD factorization
for the process in Eq.(\ref{proc}) initiated by gluons from hadrons, we only need to replace each initial  hadron  
with an on-shell gluon and to study the process: 
\begin{equation}
   g(p,a)+ g(\bar p, b) \to Q (p_1) \bar Q (p_2) + X. 
\label{P-Proc} 
\end{equation} 
In the above we have replace $h_A$ and $h_B$ with the gluon $g(p)$ and $g(\bar p)$, respectively. 
The momenta of the initial gluons are given by $p^\mu =(p^+,0,0,0)$ and $\bar p^\mu =(0,\bar p^-,0,0)$.  
The momentum of $Q$ and $\bar Q$  is given by 
\begin{equation} 
  p_1 =\frac{q}{2} + \Delta, \quad\quad p_2 =\frac{q}{2} -\Delta.
\label{MOM} 
\end{equation} 
The small velocity expansion here is an expansion in $\Delta$.    
From the $Q\bar Q$-pair one can project out a state with given quantum numbers. 
\par  
\par 
\begin{figure}[hbt]
\begin{center}
\includegraphics[width=10cm]{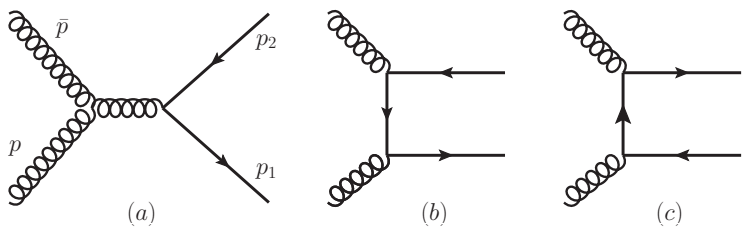}
\end{center}
\caption{Tree-level diagrams for the amplitudes of $gg \to Q\bar Q$.   }
\label{P1}
\end{figure}
\par
At tree-level, the amplitude for the partonic process in Eq.(\ref{P-Proc})  is given by diagrams in Fig.1.  
It is standard to perform the projection from the $Q\bar Q$-pair into $^3P_J^{(1)}$- and $^3 S_1^{(8)}$ state combined 
with the expansion in $\Delta$. At tree-level, we have the result for the differential 
cross-section: 
\begin{equation} 
   \frac{ d\sigma ( \chi_{0,2}) }{ d x d y d^2 q_\perp} = \frac{\pi\sigma_0 ( ^3P_{0,2} ^{(1)})}{Q^2} \delta (x y s- Q^2) \delta (1-x) \delta (1-y) 
      \delta^2 (q_\perp) ,
\label{Tree}        
\end{equation} 
with $q^+= xp^+$, $q^-=y \bar p^-$ and $s=(p+\bar p)^2$. The coefficients $\sigma_0$ are given by:  
\begin{eqnarray} 
   \sigma_0 (^3 P_0^{(1)} ) = \frac{3 (4\pi\alpha_s)^2}{N_c (N_c^2-1) m_Q^3} \langle 0 \vert {\mathcal O} (^3 P_0^{(1)}) \vert 0\rangle, \quad\quad
  \sigma_0 (^3 P_2^{(1)} ) = \frac{4 (4\pi \alpha_s)^2}{5 N_c (N_c^2-1) m_Q^3} \langle 0 \vert {\mathcal O} (^3 P_2^{(1)}) \vert 0\rangle.    
\end{eqnarray}
The matrix elements are of NRQCD operators denoted as ${\mathcal O} (^3 P_{0,2}^{(1)})$. The color-octet 
component is with the matrix element of NRQCD operator ${\mathcal O} (^3 S_{1}^{(8)})$.  The definition 
of these operators can be found in \cite{nrqcd}. 
These matrix elements characterize the transition from the produced $Q\bar Q$-pair with given quantum numbers 
into the observed quarkonium.  It is noted that at tree-level the color-octet 
component is zero.                      

\par 
\begin{figure}[hbt]
\begin{center}
\includegraphics[width=11cm]{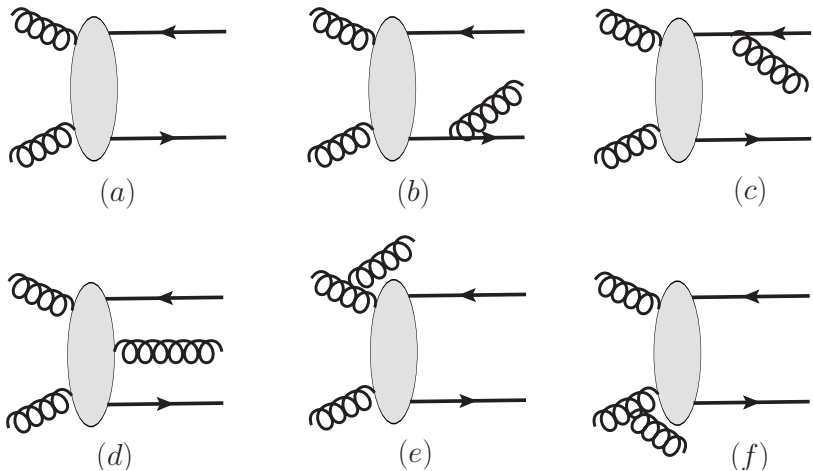}
\end{center}
\caption{Diagrams for the real correction. (a). The sum of all tree-level diagrams in Fig.1.  (d). Diagrams with one gluon emitted from propagators in tree-level diagrams. Other diagrams are for one-gluon emission from a external leg 
of tree-level diagrams. }
\label{P1}
\end{figure}
\par
\par 
Now we discuss the one-loop contribution to the process Eq.(\ref{P-Proc}). The one-loop contribution 
consists of the real- and virtual correction. In the virtual correction the unobserved state $X$ is the same 
as that in the tree-level contribution, i.e., $X$ is the vacuum. In the real correction the $X$-state 
consists of a gluon. 
The real correction is represented by diagrams in Fig.2.   
\par 
In calculating the real correction, one needs to expand it in $q_\perp/Q =\lambda$, because of  
that we are interested in the kinematical region with $\lambda \ll 1$.  In this region, the exchanged gluon 
must be collinear to the initial gluons or soft. It is straightforward 
to obtain the real contribution at the leading order of $\lambda$: 
\begin{eqnarray} 
 \frac{ d\sigma ( \chi_{0,2} )}{ d x d y d^2 q_\perp}\biggr\vert_{real} 
    &=& \frac{\pi\sigma_0 ( ^3 P_{0,2}^{(1)})}{Q^2}  \frac{ \alpha_s N_c}{\pi^2 q_\perp^2 }   \delta ( xy s- Q^2) 
\biggr [ \frac{(1-x+x^2)^2}{x (1-x)_+} \delta (1-y) 
\nonumber\\
 && + \frac{(1-y+y^2)^2}{y (1-y)_+} \delta (1-x) 
        -\delta (1-x) \delta (1-y) \ln\frac{q_\perp^2}{Q^2} \biggr ] +{\mathcal O}(\lambda^{-1}). 
\label{Real}         
\end{eqnarray}
The leading order here is at $\lambda^{-2}$ which is the same as the order of the tree-level result in Eq.(\ref{Tree}). It is singular if we take $q_\perp \to 0$. 
It is interesting to note that at one-loop only the color-singlet component 
gives the contribution at the leading order of $\lambda$. The color-octet component gives 
contributions at higher orders of $\lambda$.  
The virtual corrections from one-loop has been studied in \cite{NLOQQ}. The virtual correction for $^3 P_{J}^{(1)}$ with $J=0,2$ is:  
\begin{eqnarray} 
\frac{ d\sigma ( \chi_{J}) }{ d x d y d^2 q_\perp}\biggr\vert_{vir.} &=& \frac{\pi\sigma_0 ( ^3P_{J} ^{(1)})}{Q^2} \delta (x y s- Q^2) \delta (1-x) \delta (1-y) 
      \delta^2 (q_\perp)  \frac{\alpha_s}{\pi} \biggr [ 
  - N_c \biggr ( \frac{4}{\epsilon^2} +\frac{2}{\epsilon} \ln \frac{4\pi \mu_s^2}{e^\gamma Q^2} 
\nonumber\\  
  &&  +\frac{1}{2}  \ln^2 \frac{4\pi \mu_s^2}{e^\gamma Q^2}
        + \frac{\pi^2}{12} \biggr ) 
  - \frac{\beta_0}{2} \biggr ( \frac{2}{\epsilon} + \ln \frac{4\pi \mu_s^2}{e^\gamma Q^2} \biggr ) 
+\frac{\beta_0}{2} \ln \frac{ \mu^2}{ Q^2}  + C_J \biggr ], 
\nonumber\\
   C_0 &=&  C_F \biggr ( -\frac{7}{3} + \frac{\pi^2}{4} \biggr ) + N_c \biggr ( \frac{1}{3} + \frac{5\pi^2}{12} \biggr ), \quad 
   C_2 =  -4 C_F + N_c \biggr ( \frac{1}{3} + \frac{5}{3} \ln 2 +  \frac{\pi^2}{6} \biggr ).  
\end{eqnarray} 
The above result is obtained from the original one given in Eq.(124) of  \cite{NLOQQ} after factorizing the Coulomb singularity and subtracting the U.V. divergence. The Coulomb singularity is factorized into NRQCD matrix elements.  
The divergent terms as poles of $\epsilon=4-d$ are for collinear- or I.R. divergences. The scale $\mu_s$ is related to these divergences.  $\mu$ is the U.V. scale.                     
\par 
To study the TMD factorization of the process in Eq.(\ref{P-Proc}), one needs also to study 
the TMD gluon distribution function defined in Eq.(\ref{DEF}) by replacing $h_A$ with a free gluon. 
After the replacement, one can calculate the function with perturbative theory. The defined 
soft factor $\tilde S$ can also be calculated perturbatively.  The results at one-loop with different regularizations
of collinear- and I.R. divergences can be found in \cite{JMYG,MWZ}. 
With these results one can find that the differential cross section at one-loop accuracy can be factorized as:
\begin{eqnarray} 
\frac {d \sigma (\chi_{J})} { d x d y d^2 q_\perp} &=& \frac{\pi \sigma_0 (^3 P_{J}^{(1)})}{Q^2 } {\mathcal H}_J \int d^2k_{a\perp} d^2 k_{b\perp}
d^2 \ell_\perp  \delta^2 (\vec k_{a\perp} + \vec k_{b\perp}+\vec\ell_\perp -\vec q_\perp)  \delta (xy s -Q^2)  
\nonumber \\
   && \ \ \ \ \ \  \cdot  f_{g/A} (x,k_{a\perp},\zeta_u) f_{g/B} (y,k_{b\perp},\zeta_v) 
   \tilde S(\ell_\perp,\rho)   , 
\nonumber\\  
  {\mathcal H}_{0,2} &=& 1 +\frac{N_c \alpha_s}{4\pi} 
  \biggr [ \biggr ( \ln^2\frac{Q^2}{\zeta_u^2} + \ln^2\frac{Q^2}{\zeta_v^2} 
   - \ln\rho^2 \biggr ( 1 +  2\ln\frac{\mu^2}{Q^2} \biggr )   + 2  \ln\frac{\mu^2}{Q^2} \biggr ) 
  + \frac{4}{3}\pi^2  +6 + \frac{4}{N_c} C_{J} \biggr ] 
\nonumber\\  
   &&  +{\mathcal O}(\alpha_s),  
\label{Fac}   
\end{eqnarray} 
with the corrections suppressed by powers of $\lambda$ and the small velocity $v$. All singular contributions 
from the virtual- and real correction are factorized into TMD gluon distribution functions and 
the soft factor. Therefore, the perturbative coefficient ${\mathcal H}_{0,2}$ is finite.   The result 
shows that there is TMD factorization at one-loop for the process. In Eq.(\ref{Fac}) we have written in the factorized 
form with TMD gluon distribution functions defined in Eq.(\ref{DEF}) and the soft factor in Eq.(\ref{SoftS}).
There are different definitions of TMD parton distribution functions, e.g., the one suggested in \cite{JC1}. 
The difference between them can be calculated perturabtively and it has been studied in \cite{QQR2}. 
Taking the difference into account,  
the factorized form in Eq.(\ref{Fac}) essentially takes the same form 
as that for  different processes studied in  \cite{JMYG, SecG1,BoPi,QSW,DLPS,MWZ}. The only difference is that 
the perturbative coefficient ${\mathcal H}$ is different. The difference 
is from the difference of the definition of TMD gluon distributions and that of the considered processes. 

\par 
Since we will show the derived TMD factorization does not hold beyond one-loop, it is useful to understand 
why the factorization holds at one-loop.  In the factorization in Eq.(\ref{Fac}), the collinear divergences 
introduced by gluon-emission from initial gluons in Fig.2e and Fig.2f are factorized into TMD gluon distribution functions. 
The I.R. divergences from soft gluons emitted from initial gluons are factorized into TMD gluon distrsibution 
functions and the soft factor.  From the finiteness of the perturbative coefficient ${\mathcal H}_{0,2}$ one can realize 
that the emission of soft gluons from the $Q\bar Q$-pair in the final state does not introduce any I.R. divergent contribution, or the $Q\bar Q$ pair seems to be decoupled from soft gluons at leading power. 
Since the $Q\bar Q$ pair is in P-wave, it is in general expected that there are interactions with soft gluons. 
To completely understand this a detailed analysis of Fig.2b and 2c is needed. 
\par    
\par 
We denote the amplitude of $ g(p) + g(\bar p) \to  Q(p_1) \bar Q(p_2)$ as:
\begin{equation} 
  {\mathcal T} (Q\bar Q)  = \bar u(p_1) \Gamma (p,\bar p, p_1,p_2) v(p_2).  
\end{equation}
with $\Gamma$ is represented by the bubble in Fig.2a. The polarization vectors and color factors of initial gluons are also 
included in $\Gamma$. At tree-level, $\Gamma$ is given by the sum of all diagrams in Fig.1. 
  Now we consider the contributions from Fig.2b and 2c, in which a soft gluon with the momentum $k^{\mu} \sim Q (\lambda,\lambda,\lambda,\lambda)$ is emitted from $Q$ or $\bar Q$ in the final state
with $\lambda=q_\perp/Q\ll 1$. 
The soft gluon is with the polarization index $\mu$ and the color index $c$.  
At leading order of $\lambda$ one has the sum of the contributions as
\begin{eqnarray} 
{\mathcal T} (Q\bar Q) \biggr\vert_{2b+2c}  =  \bar u(p_1)  \biggr [ (-ig_s  T^{c} ) \frac{ i  p_1^{\mu} }{  p_1\cdot k + i\varepsilon}      
\Gamma (p,\bar p, p_1,p_2)  
 + \Gamma (p,\bar p, p_1,p_2)  \frac{ - i p_2^{\mu}  }{ p_2\cdot k + i\varepsilon} 
(-ig_s T^{c} ) \biggr ]  v(p_2), 
\end{eqnarray} 
from the above, the amplitude in general case is at order of $\lambda^{-1}$. If one calculates 
the contribution from the amplitude to the differential cross-section, one will have 
an I.R. divergent contribution when the soft gluon is not in the final state, or a  
contribution at the leading power of $\lambda$ when the soft gluon is in the final state.  These contributions need to be factorized. 
But, if we project out a state with given 
quantum numbers and make the expansion in the small velocity $v$, the order of $\lambda$ can be changed. 
\par 
For the production of $\chi_J$ we need to project the $Q\bar Q$ pair into the state which is a spin-triplet state with the orbital angular momentum $L =1$. We also need to expand the relative momentum $\Delta$ defined in Eq.(\ref{MOM}) 
and to take 
the leading order of $\Delta$ or $v$. We denote the polarization vector of the spin-triplet as $\epsilon^* (s_z)$. 
After doing the projection and the expansion we obtain the amplitude for production of a $^3P_J^{(1)}$ $Q\bar Q$ pair 
as: 
\begin{eqnarray} 
  {\mathcal T} ( ^3 P_J^{(1)} ) \biggr\vert_{2b+2c}  &\propto&  \sum_{m,s_z} \langle J,J_z\vert 
   1,m,1,s_z \rangle  \epsilon^{*\alpha} (m) \frac{\partial }{\partial \Delta ^\alpha } 
     \biggr \{  \biggr ( \frac{ g_s p_1^{\mu}}{p_1\cdot k +i\varepsilon} -  \frac{ g_s p_2^{\mu}}{p_2\cdot k +i\varepsilon} \biggr )
\nonumber\\      
     && \quad\quad\quad \frac{1}{m_Q}{\rm Tr } \biggr [  (-\gamma\cdot p_2 + m_Q) \gamma\cdot \epsilon^* (s_z) (\gamma\cdot p_1 + m_Q) T^{c} \Gamma (p,\bar p, p_1,p_2) \biggr ] \biggr \} \biggr\vert_{\Delta =0}  
\nonumber\\
   &=& \sum_{m,s_z} \langle J,J_z\vert 
   L,m,1,s_z \rangle  \epsilon^{*\alpha} (m)\biggr \{ \frac{\partial }{\partial \Delta ^\alpha } 
       \biggr ( \frac{ g_s p_1^{\mu}}{p_1\cdot k +i\varepsilon} -  \frac{ g_s p_2^{\mu}}{p_2\cdot k +i\varepsilon} \biggr ) \biggr\}
      \biggr\vert_{\Delta =0} 
\nonumber\\      
     && \quad\quad\quad \cdot {\rm Tr } \biggr ( \gamma\cdot \epsilon^* (s_z) (\gamma\cdot q + 2 m_Q) T^{c} \Gamma (p,\bar p, q/2,q/2) \biggr ) . 
\label{SoftG1}      
\end{eqnarray} 
We note that the expression of the last line is the amplitude ${\mathcal T} (^3 S_1^{(8)})$ of $g+g \to Q\bar Q(^3 S_1^{(8)})$. This amplitude is zero at tree-level.  Therefore, at one-loop the soft gluon 
is decoupled from the $^3 P_J^{(1)}$ pair at leading power. This  is why Fig.2b and Fig.2c gives no contribution 
to the one-loop real correction at leading power of $q_\perp$ as mentioned before.  The decoupling discussed 
here can be generalized to the case of emission of many soft gluons. It is noted that the soft gluon is not decoupled or its effect is not power-suppressed, if the amplitude of $g+g \to Q\bar Q(^3 S_1^{(8)})$ is nonzero. This can be checked by an explicit calculation if the soft gluon is in the final state.
One can calculate the contribution to the differential cross-section from the interference of  
the amplitude given in Eq.(\ref{SoftG1}) with the tree-level amplitude represented by Fig.2e and Fig.2f, in which the gluon 
in the final state is a soft one. 
We have for $\chi_0$: 
\begin{eqnarray} 
  \frac{ d \sigma (\chi_0)}{d x dy d^2 q_\perp}\biggr\vert_{2b+2c} =  {\mathcal F}_8  \frac{N_c\sigma_0 (^3 P_0^{(1)})}{12\pi^2 Q^2 q_\perp^2} 
 \delta (1-x) \delta (1-y) \delta (xy s-Q^2) + {\mathcal O}(\lambda^{-1}), 
\label{SoftS1}     
\end{eqnarray}    
where we have parameterized the amplitude ${\mathcal T} (^3 S_1^{(8)})$ from symmetries as:
\begin{equation} 
{\mathcal T} (^3 S_1^{(8)}) =  {\rm Tr } \biggr ( \gamma\cdot \epsilon^* (s_z) (\gamma\cdot q + 2 m_Q) T^{c} \Gamma (p,\bar p, q/2,q/2)\biggr ) = \frac{1}{m_Q} f^{abc}  \epsilon^*(s_z) \cdot( p-\bar p) \epsilon(p)\cdot \epsilon (\bar p) 
  {\mathcal F_8}.
\label{F8}   
\end{equation}  
In Eq.(\ref{F8}) the color index $a $ and the polarization $\epsilon(p)$ are of the gluon with the momentum $p$, 
the color index $b$ and the polarization $\epsilon(\bar p)$ are of the gluon with the momentum $\bar p$.
$\epsilon^*(s_z)$ has the property 
$\epsilon^*(s_z) \cdot (p+\bar p) =0$. ${\mathcal F}_8$ is a constant. 
From Eq.(\ref{SoftS1}) one can see that the soft gluon contribution is at the leading order of $\lambda$, 
i.e., the contribution is at $\lambda^{-2}$ which is the same order of the contribution given in Eq.(\ref{Real}). This clearly indicates that the effect from the soft gluon in Fig.2b and Fig.2c is not power-suppressed. 
We have also calculate the soft gluon contribution for $\chi_2$.  The contribution for $\chi_2$ is zero 
because of the conservation of angular momentum. If there are two- or more gluons in the final state, 
the contribution for $\chi_2$ can become nonzero at leading power.

\par 
If one attempts  to show the TMD factorization in Eq.(\ref{Fac}) beyond one-loop level, one needs to show 
that the decoupling holds at any order.  To show this one needs to prove that the amplitude ${\mathcal T}(^3 S_1^{(8)})$ or ${\mathcal F}_8$ is zero at any order. It is true that the amplitude 
or ${\mathcal F}_8$ is zero at tree-level by explicit 
caluclation. 
But it seems that one can not show this  from symmetries of QCD. We note here that Landau-Yang theorem 
does not apply here, because gluons, unlike photons,  have colors. By performing an one-loop calculation 
for the amplitude we find the nonzero result:
\begin{equation} 
 {\mathcal F}_8  =  \frac { \alpha_s^2}{ N_c } \left ( 17-14\ln 2 -\frac{5\pi^2}{4} \right ) +{\mathcal 
 O} (\alpha_s^3).  
\label{OCT}  
\end{equation}  
Therefore, beyond one-loop the TMD factorization in Eq.(\ref{Fac})   for the process 
in Eq.(\ref{proc},\ref{P-Proc}) does not hold, because at least there are contributions from soft gluons which are not factorized
into TMD gluon distribution functions and the defined soft factor $\tilde S$.  
\par 
At first look the factorization can be restored by modifying the soft factor for the color-singlet component 
and introducing an additional factorized contribution in Eq.(\ref{Fac}) for the color-octet contribution with 
the perturbative coefficient starting at the order of $\alpha_s^4$. The effect of soft-gluon 
emission can be completely factorized with different soft factors introduced in 
study of the resummation in heavy quark pair production in \cite{QQR2,QQR1}. E.g., 
the emission of soft gluons from the P-wave $Q\bar Q$-pair discussed for Fig.2 can be factorized
at amplitude level with 
the object built with gauge links pointing to the future: 
\begin{equation} 
   \frac{\partial }{\partial \Delta w^\alpha} \biggr (  {\mathcal L}^\dagger_{w-\Delta w} (\vec b_\perp,\infty) 
   {\mathcal L}_{w+\Delta w} (\vec b_\perp,\infty) \biggr )  \biggr\vert_{\Delta w =0}
\end{equation} 
with $w$ as the moving direction of the quarkonium. One can modify the soft factor in Eq.(\ref{SoftS})
for the color singlet component.   
However, if the factorization can be made with the modified soft factor 
in this way, it is not useful for extracting TMD gluon distribution functions, because of that we have then process-dependent soft factors which need to be determined with nonperturbative methods.  
\par
\par 
\begin{figure}[hbt]
\begin{center}
\includegraphics[width=8cm]{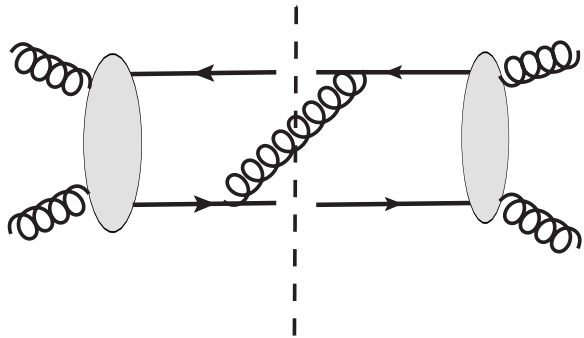}
\end{center}
\caption{ One of the four diagrams with one soft gluon emitted or absorbed by the $Q\bar Q$-pair. The soft gluon 
is in the final state and the $Q\bar Q$-pair is in color-singlet and $^3P_{0,2}$-state. The bubbles represent the amplitude for $gg\to Q\bar Q$ in color-octet and $^3 S_1$ state starting at order of $\alpha_s^2$.     }
\label{P1}
\end{figure}
\par

In fact the factorization can not be restored in the case for quarkonium production, if the color-octet component contributes at leading order of $\lambda$ and at some order of $\alpha_s$. To explain this, we consider 
the contribution to the differential cross-section from a class of diagrams given in Fig.3., where the $Q\bar Q$-pair is in color-singlet and $^3 P_{0,2}$-state. As discussed in the above, the contribution from Fig.3 with the soft gluon  
can be factorized with the modified soft factor for the color-singlet component discussed in the above. 
We note that the contribution from Fig.3 without 
the soft gluon is at leading order of $\lambda$ and factorized as the color-octet component combined with the color-octet NRQCD matrix element.      
If we integrate the momentum of the soft gluon, the contribution from Fig.3 will have an 
I.R. singularity. This I.R. divergent contribution is in fact factorized in the color-octet NRQCD matrix element of one-loop with the perturbative matching according to NRQCD factorization\cite{nrqcd,BBYL}. 
In other word, the color-octet NRQCD matrix element contains the 
same I.R. singularity. In the restored TMD factorization for the contribution from Fig.3 the momentum of the soft gluon is in fact not integrated, and the effect of the soft gluon is already factorized with the modified soft factor. 
Therefore, this I.R. singularity is double-counted. 
This implies that the I.R. singularity in the color-octet NRQCD matrix element will in turn appear 
in the perturbative coefficient and the TMD factorization can not be restored with the modified soft factor.  
This is unlike the case with the production of a free $Q\bar Q$-pair studied in \cite{QQR2,QQR1}.

\par 
Similarly, TMD factorization for the production of a spin-singlet P-wave quarkonium $h_c$ or $h_b$ denoted as $h_Q$
is already violated at one-loop level. 
According to NRQCD factorization, the differential cross-section at the leading order of $v$ is a sum of a color-singlet- and a color-octet component  
 \begin{equation} 
  d\sigma ( h_Q ) = d\sigma (^1 P_1^{(1)}) + d\sigma (^1S_0^{(8)}). 
\label{TC1} 
\end{equation}
This is similar to Eq.(\ref{TC}). In this case the color-octet component is not zero from the tree-level diagrams in Fig.1, while the color-singlet component is zero at tree-level. At one-loop with one gluon in the final state, the color-single component obtains a nonzero contribution from diagrams given 
in Fig.3 with the $Q\bar Q$-pair in color-singlet and $^1P_1$-state. Now, the bubbles in Fig.3 
stand for the amplitude $gg\to Q\bar Q$-pair in color-octet and $^1S_0$-state. The amplitude 
is nonzero at tree-level. Therefore, the $Q\bar Q$-pair is not decoupled with the soft 
gluon at one-loop level. From the study of the case with $\chi_{0,2}$, one can conclude that 
the TMD factorization for $h_Q$ does not hold at one-loop.   
We notice here that the differential cross-section for 
the production of other quarkonia with $J=1$ becomes constant for $q_\perp \to 0$. Hence, there is no 
TMD factorization. 
\par 
To summarize: We have studied TMD factorization for $P$-wave quarkonium production in hadron-hadron collisions 
at low transverse momentum. These processes are thought to be useful for extracting TMD gluon distribution 
functions of hadrons. Our study shows that the TMD factorization for the production of a quarkonium 
with $J^{PC}=0^{++}, 2^{++}$ is violated beyond one-loop level. The factorization for the production 
of $h_c$ or $h_b$ is violated already at one-loop. Therefore, 
one can not use these processes to extract TMD gluon distribution functions. To determine them from inclusive single-quarkonium production 
in hadron collisions one can only use the production of $^1S_0$ quarkonium.

\par\vskip20pt      

\noindent
{\bf Acknowledgments}
\par
We thank Dr. Y.J. Zhang for confirming the nonzero result in Eq.(\ref{OCT}) and discussions. 
The work of J.P. Ma is supported by National Nature
Science Foundation of P.R. China(No.11275244). The work of  J.X. Wang is supported by DFG and NSFC (CRC 110).
\par\vskip40pt

\par

\end{document}